\documentclass{article}%
\usepackage{amsmath}
\usepackage{amsfonts}
\usepackage{amssymb}
\usepackage{graphicx}%
\providecommand{\U}[1]{\protect\rule{.1in}{.1in}}
\providecommand{\U}[1]{\protect\rule{.1in}{.1in}}
\begin{document}

\title{Can quantum mechanics be considered consistent? a discussion of Frauchiger and
Renner's argument.}
\author{F. Lalo\"{e} \thanks{laloe@lkb.ens.fr}\\Laboratoire Kastler Brossel, ENS-Universit\'e PSL,\\CNRS, Sorbonne Universit\'e, Coll\`ege de France,\\24 rue Lhomond 75005\ Paris, France}
\date{\today}
\maketitle

\begin{abstract}
We discuss the argument proposed in Ref.~\cite{Frauchiger-Renner}, and show
that it does not particularly illustrate any inconsistency in quantum
mechanics, but rather the well known difficulty often described as the
\textquotedblleft shifty split\textquotedblright: the exact point at which the
von Neumann reduction postulate should be applied is ill defined. This is the
origin of the famous Schr\"{o}dinger's cat or Wigner's friend paradoxes.

We investigate the argument of Ref.~\cite{Frauchiger-Renner} and show that it
combines statements obtained by different agents assuming very different positions of the shifty split, and therefore applying the reduction
postulate in different ways. This results in the introduction of several different state
vectors, while such descriptions are considered as incompatible in
standard quantum mechanics. To our knowledge, no interpretation of quantum
mechanics includes this possibility; the argument thus refers to a new form of
quantum mechanics that should be specified more precisely.

\end{abstract}

\bigskip

In Ref. \cite{Frauchiger-Renner}, the authors propose a generalization of the
\textquotedblleft Wigner friend paradox\textquotedblright\ with a thought
experiment illustrating the possible existence of inconsistencies in quantum
mechanics. The purpose of the present note is to study their argument in more
detail and, in particular, to write all relevant state vectors explicitly.
This provides a useful guide to obtain a clearer view of the conclusions that
can be drawn from the argument.

A well-known fundamental difficulty of quantum mechanics is that the
fundamental equation of its dynamics (the Schr\"{o}dinger equation giving the
evolution of the state vector) allows the appearance of superpositions of
macroscopically distinct states; since these superpositions are not observed,
one needs to introduce a limit to the validity of the equation\footnote{Unless
one gives up macroscopic uniqueness, as in the Everett interpretation, where
the difficulties are different.}. This difficulty is illustrated by the famous
Schr\"{o}dinger cat paradox \cite{Schrodinger-cat} where, at the end of the
scenario, the cat reaches a superposition state that is obviously meaningless
(a \textquotedblleft ridiculous case\textquotedblright\ in the words of
Schr\"{o}dinger in 1935).\ Another example is the \textquotedblleft Wigner
friend paradox\textquotedblright\ introduced by\ Wigner in 1961
\cite{Wigner-friend}, where a friend and his whole laboratory are supposed to
reach a superposition of macroscopically distinct states -- another
illustration of the fact that the linear Schr\"{o}dinger dynamics cannot be
obeyed too far.

Actually, this problem was already well identified by von Neumann long before,
in 1932 \cite{Von-Neumann}. Treating the whole measurement process quantum
mechanically, he pointed out that the consistency of the theory requires that
measurement should not be treated with the usual deterministic dynamics.\ He
therefore introduced the \textquotedblleft projection
postulate\textquotedblright\ which, under certain conditions, attributes a
second rule for the evolution to the state vector. This solves the problem,
but a real difficulty subsists: since the borders between the two different
dynamics of the state vector are not precisely specified, the theory becomes
ill defined.\ No-one knows exactly where to put this border, and physicists
must resort to rules that are valid only FAPP (for all practical
purposes).\ In the abundant literature on the subject, it has often be named
as the \textquotedblleft shifty split\textquotedblright\ problem.\ We note in
passing that the subject is not purely academic: the present progress of
experimental techniques is such that the borders between the microscopic and
macroscopic world are becoming more and more accessible.

Since the very essence of the Wigner paradox lies in possible contradictory
evolutions of the state vector, we will complete the argument of
\cite{Frauchiger-Renner}\ with calculations of the state vector in various
situations - this reference evokes the calculation but never really completes
it. Physics makes predictions with the help of equations, and in this case the
equations contain the state vector (or the density operator); studying its
exact expression is therefore necessary, independently of considerations on
the interpretation of the theory.\ Since the question is to decide if and
where state vector reduction should be applied, we will not make a selection
between the various possible choices: we will perform the calculations in
various scenarios, applying the reduction at different steps of the
experiment, and trying in this way to exhaust the various possibilities to apply quantum mechanics consistently.\ As expected, all these different ways to apply quantum mechanics
lead to different predictions, but we will check that none of these
calculations leads to any inconsistency.

To facilitate the discussion, we use exactly the same notation as Ref.
\cite{Frauchiger-Renner}\ and, for the sake of simplicity, we ignore the time
evolution between the different measurements of the protocol described by
these authors. This amounts to assuming that the whole experiment lasts a
short time. Nevertheless, if this is not the case, it is straightforward to
add unitary evolution operators wherever needed; this does not change the
structure of the argument and its conclusions.

\section{A first experiment}

\label{first-experiment}

In \S~\ref{first-expt-no-projn} we first review the successive steps of the
protocol and calculate the  evolution of the state vector,
without applying any von Neumann projection (this can be seen as the Everett
calculation). To be sure that no entanglement effect has been missed, we write
the state of all systems and all agents explicitly, even if this may lead to a
somewhat cumbersome notation. The reader who is not particularly interested in
these technicalities is invited to skip them and to go directly to the
discussions of \S\S ~\ref{premiere-discussion} and \ref{discussion}.

Without state vector projection, the expression of the final state vector
shows that all the statements of table II of Ref. \cite{Frauchiger-Renner} are
not recovered together. We will then try to apply state vector projection
and, since standard quantum mechanics is not very specific about when it
should be applied, we will try different possibilities.\ In
\S \ \ref{first-expt-F-projn} we examine the case where agent $\overline{F}$
projects the state vector, in \S~\ref{F-andFbar-proj} where two agents
project it, and in \S~\ref{first-expt-every-agent} the case where all agents
project the state vector.

\subsection{No projection of the state vector}
\label{first-expt-no-projn}

We begin by a standard calculation including no state vector projection at all.

\subsubsection{Successive expressions of the state vector}

The first measurement is performed by an agent $\overline{F}$ (the letter F is
for \textquotedblleft friend\textquotedblright) with a qubit that is initially
in a coherent superposition:%
\begin{equation}
\left\vert \Psi\right\rangle =\frac{1}{\sqrt{3}}\left\vert \text{heads}%
\right\rangle +\sqrt{\frac{2}{3}}\left\vert \text{tails}\right\rangle
\label{0}%
\end{equation}
If $\left\vert \overline{F}:h\right\rangle $ is the state of $\overline{F}$
(and her laboratory $\overline{L}$) if she has observed a heads, and
$\left\vert \overline{F}:t\right\rangle $ the state if she has observed a
tails, after the first measurement the state of the whole system is:%
\begin{equation}
\left\vert \Psi_{1}\right\rangle =\frac{1}{\sqrt{3}}\left\vert \text{heads}%
\right\rangle \left\vert \overline{F}:h\right\rangle +\sqrt{\frac{2}{3}%
}\left\vert \text{tails}\right\rangle \left\vert \overline{F}:t\right\rangle
\label{1}%
\end{equation}
According to the scenario of \cite{Frauchiger-Renner}, $\overline{F}$ then
sends a spin to $F$ in a state that depends on her result of measurement: a
state $\left\vert S:\downarrow\right\rangle $ down polarized along axis $Oz$
if she observes a heads, a state polarized in the orthogonal direction $Ox$ is
she observes a tails. This leads to the state:%
\begin{align}
\left\vert \Psi_{2}\right\rangle  &  =\frac{1}{\sqrt{3}}\left\vert
\text{heads}\right\rangle \left\vert \overline{F}:h\right\rangle \left\vert
S:\downarrow\right\rangle +\sqrt{\frac{2}{3}}\left\vert \text{tails}%
\right\rangle \left\vert \overline{F}:t\right\rangle ~\left[  \frac{1}%
{\sqrt{2}}\left\vert S:\downarrow\right\rangle +\frac{1}{\sqrt{2}}\left\vert
S:\uparrow\right\rangle \right] \nonumber\\
&  =\frac{1}{\sqrt{3}}\Big[\left\vert \text{heads}\right\rangle \left\vert
\overline{F}:h\right\rangle +\left\vert \text{tails}\right\rangle \left\vert
\overline{F}:t\right\rangle \Big]\left\vert S:\downarrow\right\rangle
+\frac{1}{\sqrt{3}}\left\vert \text{tails}\right\rangle \left\vert
\overline{F}:t\right\rangle \left\vert S:\uparrow\right\rangle \label{2}%
\end{align}
We now assume that the second friend $F$ performs a measurement of the $S_{z}$
component of the spin; $\left\vert F:\uparrow\right\rangle $ and $\left\vert
F:\downarrow\right\rangle $ are the states reached by $F$ (and her laboratory
$L$) corresponding to the two results of measurement. The state vector then
becomes:%
\begin{equation}
\left\vert \Psi_{3}\right\rangle =\frac{1}{\sqrt{3}}\Big[\left\vert
\text{heads}\right\rangle \left\vert \overline{F}:h\right\rangle +\left\vert
\text{tails}\right\rangle \left\vert \overline{F}:t\right\rangle
\Big]\left\vert S:\downarrow\right\rangle \left\vert F:\downarrow\right\rangle
+\frac{1}{\sqrt{3}}\left\vert \text{tails}\right\rangle \left\vert
\overline{F}:t\right\rangle \left\vert S:\uparrow\right\rangle \left\vert
F:\uparrow\right\rangle \label{3}%
\end{equation}

At this stage, Ref. \cite{Frauchiger-Renner} introduces an external operator
$\overline{W}$ (the letter W emphasizes that this operator plays Wigner's role
in the usual paradox) who performs a measurement on $\overline{L}$ with
eigenstates\footnote{Of course, assuming the existence of theses states is not
usual, since they are coherent superpositions of macroscopically distinct
states (states where $\overline{F}$ has registered different results in her
memory register). Creating a pure state involving a coherent superposition of
entire laboratories, such as those written below in (\ref{5}), seems totally
irrealistic. In fact, one generally assumes that such states do not exist,
which means that the measurement procedure in question is impossible.\ But it
is precisely the point of Wigner's friend paradox and of
Ref.\ \cite{Frauchiger-Renner} to discuss the possible effects of such macroscopic
superpositions, and to introduce pure thought experiments; here we follow the
same approach.} associated respectively to results \textquotedblleft%
$\overline{\text{OK}}$\textquotedblright\ and \textquotedblleft$\overline
{\text{fail}}$\textquotedblright:%
\begin{align}
\left\vert \overline{\text{OK}}\right\rangle _{\overline{L}}  &  =\frac
{1}{\sqrt{2}}\Big[\left\vert \text{heads}\right\rangle \left\vert \overline
{F}:h\right\rangle -\left\vert \text{tails}\right\rangle \left\vert
\overline{F}:t\right\rangle \Big]\nonumber\\
\left\vert \overline{\text{fail}}\right\rangle _{\overline{L}}  &  =\frac
{1}{\sqrt{2}}\Big[\left\vert \text{heads}\right\rangle \left\vert \overline
{F}:h\right\rangle +\left\vert \text{tails}\right\rangle \left\vert
\overline{F}:t\right\rangle \Big] \label{4}%
\end{align}
As remarked by Bub \cite{Bub-2017}, the ket $\left\vert \Psi_{3}\right\rangle
$ has three other equivalent expressions:%
\begin{align}
\left\vert \Psi_{3}\right\rangle  &  =\frac{1}{\sqrt{3}}\Big[\left\vert
\text{heads}\right\rangle \left\vert \overline{F}:h\right\rangle \left\vert
S:\downarrow\right\rangle \left\vert F:\downarrow\right\rangle +\left\vert
\text{tails}\right\rangle \left\vert \overline{F}:t\right\rangle \left\vert
S:\downarrow\right\rangle \left\vert F:\downarrow\right\rangle +\left\vert
\text{tails}\right\rangle \left\vert \overline{F}:t\right\rangle \left\vert
S:\uparrow\right\rangle \left\vert F:\uparrow\right\rangle \Big]\nonumber\\
&  =\sqrt{\frac{2}{3}}\left\vert \overline{\text{fail}}\right\rangle
_{\overline{L}}\left\vert S:\downarrow\right\rangle \left\vert F:\downarrow
\right\rangle +\frac{1}{\sqrt{3}}\left\vert \text{tails}\right\rangle
\left\vert \overline{F}:t\right\rangle \left\vert S:\uparrow\right\rangle
\left\vert F:\uparrow\right\rangle \nonumber\\
&  =\frac{1}{\sqrt{3}}\left\vert \text{heads}\right\rangle \left\vert
\overline{F}:h\right\rangle \left\vert S:\downarrow\right\rangle \left\vert
F:\downarrow\right\rangle +\sqrt{\frac{2}{3}}\left\vert \text{tails}%
\right\rangle \left\vert \overline{F}:t\right\rangle \left\vert \text{fail}%
\right\rangle _{L} \label{6-ter}%
\end{align}
(the ket $\left\vert \text{fail}\right\rangle _{L}$ in the third line will be
defined below, in (\ref{6})). After $\overline{W}$'s measurement, the state
vector written in (\ref{3}) becomes:
\begin{equation}
\left\vert \Psi_{4}\right\rangle =\sqrt{\frac{2}{3}}\left\vert \overline
{\text{fail}}\right\rangle _{\overline{L}}\left\vert \overline{W}%
:\overline{\text{fail}}\right\rangle \left\vert S:\downarrow\right\rangle
\left\vert F:\downarrow\right\rangle +\frac{1}{\sqrt{6}}\Big[\left\vert
\overline{\text{fail}}\right\rangle _{\overline{L}}\left\vert \overline
{W}:\overline{\text{fail}}\right\rangle -\left\vert \overline{\text{OK}%
}\right\rangle _{\overline{L}}\left\vert \overline{W}:\overline{\text{OK}%
}\right\rangle \Big]\left\vert S:\uparrow\right\rangle \left\vert
F:\uparrow\right\rangle \label{5}%
\end{equation}
where $\left\vert \overline{W}:\overline{\text{OK}}\right\rangle $ (and
$\left\vert \overline{W}:\overline{\text{fail}}\right\rangle $) are the states
describing $\overline{W}$ (and his laboratory) having registered result
\ $\overline{\text{OK}}$ (and $\overline{\text{fail}}$) respectively.\ At this
stage, we check that\ $\left\vert \overline{W}:\overline{\text{OK}%
}\right\rangle $ is perfectly correlated with $\left\vert S:\uparrow
\right\rangle $, which shows that the third statement $s_{Q}^{\overline{W}}$ of Table II is correct.

This perfect correlation occurs because, in (\ref{3}), the component of $\left\vert \Psi_{3}\right\rangle$ on a down spin state is exactly orthogonal to $\left\vert \overline{\text{OK}}\right\rangle _{\overline{L}}$; this is a destructive interference effect between the $\left\vert \text{heads}\right\rangle$ and $\left\vert \text{tails}\right\rangle$ components of $\left\vert \Psi_{3}\right\rangle$ (opposite results obtained by $\overline F$). Of course, if the $\left\vert \text{heads}\right\rangle$ component of (\ref{3}) was removed (because $\overline{F}$ has observed result \textquotedblleft tails\textquotedblright ), this effect would non longer occur and the correlation would vanish.

Finally, the external operator $W$ performs a measurement on $L$ with
eigenstates:%
\begin{align}
\left\vert \text{OK}\right\rangle _{L}  &  =\frac{1}{\sqrt{2}}\Big[\left\vert
S:\downarrow\right\rangle \left\vert F:\downarrow\right\rangle -\left\vert
S:\uparrow\right\rangle \left\vert F:\uparrow\right\rangle \Big]\nonumber\\
\left\vert \text{fail}\right\rangle _{L}  &  =\frac{1}{\sqrt{2}}%
\Big[\left\vert S:\downarrow\right\rangle \left\vert F:\downarrow\right\rangle
+\left\vert S:\uparrow\right\rangle \left\vert F:\uparrow\right\rangle
\Big] \label{6}%
\end{align}
In the corresponding basis, $\left\vert \Psi_{4}\right\rangle $ becomes:%
\begin{align}
\left\vert \Psi_{4}\right\rangle  &  =\frac{1}{\sqrt{12}}\left\vert
\overline{\text{OK}}\right\rangle _{\overline{L}}\left\vert \overline
{W}:\overline{\text{OK}}\right\rangle \left\vert \text{OK}\right\rangle
_{L}-\frac{1}{\sqrt{12}}\left\vert \overline{\text{OK}}\right\rangle
_{\overline{L}}\left\vert \overline{W}:\overline{\text{OK}}\right\rangle
_{\overline{L}}\left\vert \text{fail}\right\rangle _{L}\nonumber\\
&  +\frac{1}{\sqrt{12}}\left\vert \overline{\text{fail}}\right\rangle
_{\overline{L}}\left\vert \overline{W}:\text{fail}\right\rangle \left\vert
\text{OK}\right\rangle _{L}+\frac{\sqrt{3}}{2}\left\vert \overline
{\text{fail}}\right\rangle _{\overline{L}}\left\vert \overline{W}%
:\text{fail}\right\rangle \left\vert \text{fail}\right\rangle _{L}
\label{6-bis}%
\end{align}
If $\left\vert W:\text{OK}\right\rangle $ and $\left\vert W:\text{fail}%
\right\rangle $ are the states describing $W$ (and his laboratory) after this
measurement, the state vector finally becomes:%
\begin{align}
\left\vert \Psi_{5}\right\rangle  &  =\frac{1}{\sqrt{12}}\left\vert
\overline{\text{OK}}\right\rangle _{\overline{L}}\left\vert \overline
{W}:\overline{\text{OK}}\right\rangle \left\vert \text{OK}\right\rangle
_{L}\left\vert W:\text{OK}\right\rangle -\frac{1}{\sqrt{12}}\left\vert
\overline{\text{OK}}\right\rangle _{\overline{L}}\left\vert \overline
{W}:\overline{\text{OK}}\right\rangle \left\vert \text{fail}%
\right\rangle _{L}\left\vert W:\text{fail}\right\rangle \nonumber\\
&  +\frac{1}{\sqrt{12}}\left\vert \overline{\text{fail}}\right\rangle
_{\overline{L}}\left\vert \overline{W}:\text{fail}\right\rangle \left\vert
\text{OK}\right\rangle _{L}\left\vert W:\text{OK}\right\rangle +\frac{\sqrt
{3}}{2}\left\vert \overline{\text{fail}}\right\rangle _{\overline{L}%
}\left\vert \overline{W}:\text{fail}\right\rangle \left\vert \text{fail}%
\right\rangle _{L}\left\vert W:\text{fail}\right\rangle \label{7}%
\end{align}
Relation (\ref{7}) provides what could be called the \textquotedblleft Everett
view\textquotedblright\ of the experiment, where no state vector projection
ever occurs and all observers are entangled. In particular we remark that,
because of $W$'s entanglement, the state has now acquired a non-zero component over $\left\vert
\overline{\text{OK}}\right\rangle _{\overline{L}}\left\vert \overline
{W}:\overline{\text{OK}}\right\rangle\left\vert S:\downarrow\right\rangle \left\vert F:\downarrow\right\rangle$: if we insert (\ref{6}) into (\ref{7}) instead of (\ref{6-bis}), the component no longer vanishes. Therefore $\overline {W}$ observing $\overline{\text{OK}}$ does not mean that the spin is necessarily up.

\subsubsection{Correlations between measurement results}

\label{correlations}

We first notice on (\ref{7}) that, one times out of $12$, the two results
$\overline{\text{OK}}$ and OK are obtained, in agreement with the fourth
statement $s_{Q}^{W}$ of Table II of Ref. \cite{Frauchiger-Renner}.

($i$) The r=tails component of (\ref{7}) is obtained by using relations (\ref{4}):%
\begin{align}
\left\vert \Psi_{5}\right\rangle _{\text{tails}}  &  =\left\vert
\text{tails}\right\rangle \left\vert \overline{F}:t\right\rangle \left[
-\frac{1}{\sqrt{12}}\left\vert \overline{W}:\overline{\text{OK}}\right\rangle
\left\vert \text{OK}\right\rangle _{L}\left\vert W:\text{OK}\right\rangle
+\frac{1}{\sqrt{12}}\left\vert \overline{W}:\overline{\text{OK}}\right\rangle
_{\overline{L}}\left\vert \text{fail}\right\rangle _{L}\left\vert
W:\text{fail}\right\rangle \right. \nonumber\\
&  ~~~~~~~~~~\left.  +\frac{1}{\sqrt{12}}\left\vert \overline{W}%
:\text{fail}\right\rangle \left\vert \text{OK}\right\rangle _{L}\left\vert
W:\text{OK}\right\rangle +\frac{\sqrt{3}}{2}\left\vert \overline
{W}:\text{fail}\right\rangle \left\vert \text{fail}\right\rangle
_{L}\left\vert W:\text{fail}\right\rangle \right]  \label{9-5}%
\end{align}
which shows that results r=tails and OK are compatible; the first statement
$s_{Q}^{\overline{F}}$ of Table II\ of Ref. \cite{Frauchiger-Renner} is not valid.

If, nevertheless, one suppresses the kets describing the state of
$\overline{W}$ from (\ref{9-5}), the first and the third term in the right hand
side of this expression cancel each other; the component on $\left\vert
W:\text{OK}\right\rangle $ disappears by a destructive interference effect -- exactly as in the third line of (\ref{6-ter}),
where $\left\vert \text{tails}\right\rangle $ is associated with $\left\vert
\text{fail}\right\rangle _{L}$ only.\ This (arbitrary) operation restores the first statement of Table II. But, since
$W$ is actually entangled with $\overline{F}$'s laboratory in the full state vector (\ref{9-5}), this destructive
interference effect does not occur.

($ii$) The
r=heads component is obtained in the same way as (\ref{9-5}); it also contains 4 orthogonal
components, which restores the $\left\vert \text{heads}\right\rangle
\left\vert \overline{F}:h\right\rangle \left\vert S:\uparrow\right\rangle
\left\vert F:\uparrow\right\rangle $ component that had disappeared from the
first line of (\ref{6-ter}). From this discussion we conclude that, if one
wishes to study the correlations between the results of $\overline{F}$ and
$W$, one must take into account the perturbation introduced by $\overline{W}%
$'s measurement: $\overline{F}$'s initial statement is valid only if she
ignores the future perturbations introduced by this measurement.

($iii$) We can also start from (\ref{7}) and expand the ket $\left\vert \overline{\text{OK}}(\overline
{L},\overline{W})\right\rangle \left\vert \text{OK}(L,W)\right\rangle $ by
inserting the definitions (\ref{4}) and (\ref{6}) of the kets in the product;
we then obtain coherent superpositions of two possible states of the coin, of
the two operators $\overline{F}$ and $F$, and of the spin:%
\begin{align}
\left\vert \overline{\text{OK}}(\overline{L},\overline{W})\right\rangle
\left\vert \text{OK}(L,W)\right\rangle  &  =\frac{1}{\sqrt{2}}\Big[\left\vert
\text{heads}\right\rangle \left\vert \overline{F}:h\right\rangle -\left\vert
\text{tails}\right\rangle \left\vert \overline{F}:t\right\rangle
\Big]\left\vert \text{OK}(L,W)\right\rangle \nonumber\\
&  =\frac{1}{\sqrt{2}}\left\vert \overline{\text{OK}}(\overline{L}%
,\overline{W})\right\rangle \Big[\left\vert S:\downarrow\right\rangle
\left\vert F:\downarrow\right\rangle -\left\vert S:\uparrow\right\rangle
\left\vert F:\uparrow\right\rangle \Big] \label{9-2}%
\end{align}
The first line shows again that result OK is possible after result
r=tails.\ The second line shows that the spin can be in any state if
$\overline{OK}$ is obtained, in contradiction with statement $s_{Q}%
^{\overline{W}}$. The same is true for other combinations of results obtained
by $\overline{W}$ and $W$.

($iv$) The conclusion of this subsection it that, in terms of assumption (S) of Ref. \cite{Frauchiger-Renner}, the
average value of the projector $\pi_{\xi}^{H}$ considered in Assumption (Q) is
not equal to unity. It would certainly be the case only if the perturbations
introduced by $\overline{W}$ at an intermediate time were ignored, but this
would have no justification: in quantum mechanics, measurements do introduce
perturbations. As we see in the next subsection, recovering statement
$s_{Q}^{\overline{F}}$ is indeed possible, but requires that $\overline{F}$
should apply the postulate of state vector reduction, which then changes other predictions.

\subsection{Operator $\overline{F}$ projects the state vector}
\label{first-expt-F-projn}

We can try to improve the correlations by using state vector projection. If operator
$\overline{F}$ projects the state vector under the effect of her measurement,
and if she observes \textquotedblleft tails\textquotedblright, relation
(\ref{1}) then becomes:%
\begin{equation}
\left\vert \Psi_{1}^{\prime}\right\rangle =\left\vert \text{tails}%
\right\rangle \left\vert \overline{F}:t\right\rangle \label{1-prime}%
\end{equation}
and (\ref{2}) is replaced by:%
\begin{equation}
\left\vert \Psi_{2}^{\prime}\right\rangle =\left\vert \text{tails}%
\right\rangle \left\vert \overline{F}:t\right\rangle ~\left[  \frac{1}%
{\sqrt{2}}\left\vert S:\downarrow\right\rangle +\frac{1}{\sqrt{2}}\left\vert
S:\uparrow\right\rangle \right]  \label{2-prime}%
\end{equation}
The state vector just after the second measurement by $F$ now becomes, instead
of (\ref{3}) :%
\begin{equation}
\left\vert \Psi_{3}^{\prime}\right\rangle =\left\vert \text{tails}%
\right\rangle \left\vert \overline{F}:t\right\rangle ~\left[  \frac{1}%
{\sqrt{2}}\left\vert S:\downarrow\right\rangle \left\vert F:\downarrow
\right\rangle +\frac{1}{\sqrt{2}}\left\vert S:\uparrow\right\rangle \left\vert
F:\uparrow\right\rangle \right]  \label{3-prime}%
\end{equation}
which, in the basis of the eigenstates (\ref{4}) performed by $\overline{W}$
is nothing but:%
\begin{equation}
\text{ }\left\vert \Psi_{3}^{\prime}\right\rangle =\frac{1}{2}\Big[\left\vert
\overline{\text{fail}}\right\rangle _{\overline{L}}-\left\vert \overline
{\text{OK}}\right\rangle _{\overline{L}}\Big]~\Big[\left\vert S:\downarrow
\right\rangle \left\vert F:\downarrow\right\rangle +\left\vert S:\uparrow
\right\rangle \left\vert F:\uparrow\right\rangle \Big] \label{4-prime}%
\end{equation}

($i$) After $\overline{W}$'s measurement, this state becomes:%
\begin{equation}
\text{ }\left\vert \Psi_{4}^{\prime}\right\rangle =\frac{1}{2}\Big[\left\vert
\overline{\text{fail}}\right\rangle _{\overline{L}}\left\vert \overline
{W}:\overline{\text{fail}}\right\rangle -\left\vert \overline{\text{OK}%
}\right\rangle _{\overline{L}}\left\vert \overline{W}:\overline{\text{OK}%
}\right\rangle \Big]~\Big[\left\vert S:\downarrow\right\rangle \left\vert
F:\downarrow\right\rangle +\left\vert S:\uparrow\right\rangle \left\vert
F:\uparrow\right\rangle \Big] \label{4-prime-bis}%
\end{equation}
which, in the basis if the eigenstates (\ref{6}) performed by $W$, reads:%
\begin{equation}
\left\vert \Psi_{4}^{\prime}\right\rangle =\frac{1}{\sqrt{2}}\Big[\left\vert
\overline{\text{fail}}\right\rangle _{\overline{L}}\left\vert \overline
{W}:\overline{\text{fail}}\right\rangle -\left\vert \overline{\text{OK}%
}\right\rangle _{\overline{L}}\left\vert \overline{W}:\overline{\text{OK}%
}\right\rangle \Big]~\left\vert \text{fail}\right\rangle _{L} \label{5-prime}%
\end{equation}

We then see that result OK is never obtained. Physically this is because, if
result \textquotedblleft tails\textquotedblright\ only is selected,
$\overline{F}$ sends to $L$ a transversely polarized spin, which after
measurement by $F$ projects $F$ and $L$ into a state that is orthogonal to
$\left\vert \text{OK}\right\rangle _{L}$; a destructive quantum interference
effect then forbids result OK.\ Moreover, $\overline{W}$'s measurement
performed on laboratory $\overline{L}$ does not change\footnote{Agents
$\overline{W}$ and $W$ perform their measurements on two different systems,
$\overline{L}$ and $L$, which can be at an arbitrary distance.\ Quantum
mechanics does not allow superluminal communication, and it obeys the so
called NS conditions \cite{Barrett-Linden-et-al, Masanes-et-al-2006}.\ What is
nevertheless important is the order in which measurements are performed on one
site, since the corresponding projectors do not commute. In
\S~\ref{reversing}, we study what happens if the order of measurements is
reversed.} the state of $L$, so that it cannot reintroduce result OK.\ Result
\textquotedblleft fail\textquotedblright\ is therefore always obtained by $W$,
and the first statement $s_{Q}^{\overline{F}}$ in Table II of Ref.
\cite{Frauchiger-Renner} is now correct.

($ii$) Nevertheless, the third statement $s_{Q}^{\overline{W}}$ of Table II is now not correct: in (\ref{4-prime}), the correlation between results associated with
$\left\vert \overline{W}:\overline{\text{OK}}\right\rangle $ and $\left\vert
S:\uparrow\right\rangle $, which was perfect in (\ref{5}), has disappeared.  Therefore, $\overline{W}$ obtaining result $\overline{\text{OK}}$ no longer means that $W$ can sometimes also obtain result $OK$.

This is because
the initial projection of the state vector performed after  $\overline{F}$'s measurement has destroyed a $\left\vert
\text{heads}\right\rangle$ component of the sate vector that interfered destructively with another component, and was the origin of the  perfect correlation. To understand how a second component can reappear after the first one has been cancelled, we can  expand the bracket of (\ref{5-prime}) by inserting relations
(\ref{4}) and obtain:%
\begin{align}
\left\vert \Psi_{4}^{\prime}\right\rangle  &  =\frac{1}{2}\Big[\left( \vphantom{\frac{1}{2}}
\left\vert \text{heads}\right\rangle \left\vert \overline{F}:h\right\rangle
+\left\vert \text{tails}\right\rangle \left\vert \overline{F}:t\right\rangle
\right)  \left\vert \overline{W}:\overline{\text{fail}}\right\rangle
\nonumber\\
&  ~~~~~~~~~~~~~~-\left( \vphantom{\frac{1}{2}} \left\vert \text{heads}\right\rangle \left\vert
\overline{F}:h\right\rangle -\left\vert \text{tails}\right\rangle \left\vert
\overline{F}:t\right\rangle \right)  \left\vert \overline{W}:\overline
{\text{OK}}\right\rangle \Big]~\left\vert \text{fail}\right\rangle _{L}
\label{5-ter}%
\end{align}
which indeed has a non-zero component on $\left\vert
\text{heads}\right\rangle$, even if this is not the case in the previous states $\left\vert \Psi_{2}^{\prime}\right\rangle $ and $\left\vert \Psi_{3}^{\prime}\right\rangle $. The properties of (\ref{5-ter}) are similar to those of (\ref{9-5}): the component
result $\left\vert
\text{heads}\right\rangle$ vanishes if the state of $\overline{W}$ is
ignored (the entanglement with $\overline{W}$ is ignored), but no longer does when entanglement is included.

This shows that, as
above, it is the perturbing effect of  $\overline{W}$'s measurement on system
$\overline{L}$ that changes the correlations; here, it reintroduces a
component on $\left\vert
\text{heads}\right\rangle$ into a ket that had no such component. In other words, agent $\overline{W}$ changes the content of the memory of $\overline{F}$, who thought that she has observed r=tails, but once perturbed no longer remembers any specific result.

\subsection{$\overline{F}$ and $F$ project the state vector}
\label{F-andFbar-proj}

If $F$ observes $\left\vert S:\uparrow\right\rangle $, relation (\ref{3-prime})
is replaced by:%
\begin{align}
\left\vert \Psi_{2}^{\prime}\right\rangle  &  =\left\vert \text{tails}%
\right\rangle \left\vert \overline{F}:t\right\rangle \left\vert S:\uparrow
\right\rangle \left\vert F:\uparrow\right\rangle =\frac{1}{\sqrt{2}%
}\Big[\left\vert \overline{\text{fail}}\right\rangle _{\overline{L}}\left\vert
\overline{W}:\overline{\text{fail}}\right\rangle -\left\vert \overline
{\text{OK}}\right\rangle _{\overline{L}}\left\vert \overline{W}:\overline
{\text{OK}}\right\rangle \Big]\left\vert S:\uparrow\right\rangle \left\vert
F:\uparrow\right\rangle \nonumber\\
&  =\frac{1}{2}\Big[\left\vert \overline{\text{fail}}\right\rangle
_{\overline{L}}-\left\vert \overline{\text{OK}}\right\rangle _{\overline{L}%
}\Big]\Big[\left\vert \overline{\text{fail}}\right\rangle _{\overline{L}%
}-\left\vert \text{OK}\right\rangle _{\overline{L}}\Big] \label{201}%
\end{align}
If $F$ observes $\left\vert S:\downarrow\right\rangle $, the only difference
is that a plus sign replaces the minus sign in the second bracket. In any
case, the state is a product: whatever $\overline{W}$'s result is, no
information is obtained on the spin.\ Actually, neither the first statement
$s_{Q}^{\overline{F}}$ nor the third $s_{Q}^{\overline{W}}$ of Table II are
then correct. Nevertheless, $F$'s measurement has changed the state of his
laboratory $L$ and restored the possibility of $W$ observing result OK.

\subsection{Every agent projects the state vector}
\label{first-expt-every-agent}

We now assume that each of the four agents projects the state vector.\ The
last two measurements, by $W$ and $W$, detemine the final form of the state
vector, which is nothing but one of the four components of (\ref{7}).\ If, for
instance, these two agents obtained results $\overline{\text{OK}}$  and
OK,\ we retain only the first component, and obtain the final state vector
(after normalization) as:%
\begin{align}
\left\vert \Psi_{5}^{\prime\prime}\right\rangle  & =\left\vert \overline
{\text{OK}}\right\rangle _{\overline{L}}\left\vert \text{OK}\right\rangle
_{L}\left\vert \overline{W}:\overline{\text{OK}}\right\rangle \left\vert
W:\text{OK}\right\rangle \nonumber\\
& =\frac{1}{2}\Big[\left\vert \text{heads}\right\rangle \left\vert
\overline{F}:h\right\rangle -\left\vert \text{tails}\right\rangle \left\vert
\overline{F}:t\right\rangle \Big]\Big[\left\vert S:\downarrow\right\rangle
\left\vert F:\downarrow\right\rangle -\left\vert S:\uparrow\right\rangle
\left\vert F:\uparrow\right\rangle \Big]\left\vert \overline{W}:\overline
{\text{OK}}\right\rangle \left\vert W:\text{OK}\right\rangle \label{7-prime}%
\end{align}
The observation of result $\overline{\text{OK}}$ provides no information on
the present state of the spin; the observation of result OK is not
incompatible with tails.

\subsection{Discussion}
\label{premiere-discussion}

Assume now that a \textquotedblleft superoperator\textquotedblright\ O decides
to build a logical reasoning by combining the statements made by the various
agents who perform measurements, including of course the two friends
$\overline{F}$ and $F$.\ The superoperator attempts to find a logical framework in which he can obtain all lines of Table II of Ref.~\cite{Frauchiger-Renner}. For this purpose, he can take different points of view,
depending whether he considers that $\overline{F}$ and $F$ (and their laboratories), just after they perform
a measurement, should necessarily be described by quantum states associated with
one single result (as one usually does in quantum mechanics), or can also
be described by superposition of such states (as in Wigner's friend original
paradox). Since several agents play a role in the experiment, different
choices are possible.

(i) The most natural point of view is to consider that that $\overline{F}$ can communicate valid
statements about her results to O only when she and her laboratory (position of the pointer of her apparatus, etc.) are described by an eigenstate
associated with her measurement, for instance $\left\vert \text{tails}%
\right\rangle $. Otherwise her \textquotedblleft memory register\textquotedblright\ (in\ Everett's terms)
is in a state containing at the same time different results; under these
conditions, one may wonder how she could express any statement with certainty
to O. But then, since $\overline{F}$ is not in a
quantum superposition when $\overline{W}$ starts his measurement, we have seen in
\S \ \ref{first-expt-F-projn} that result
$\overline{\text{OK}}$ does not necessarily implies that the spin is in state
$\left\vert S:\uparrow\right\rangle $; statement $s_Q^{\overline W}$ is not correct.

A similar difficulty occurs with $s_Q^W$ statement:
after  $\overline{F}$ has obtained \textquotedblleft tails\textquotedblright , we have seen
that she can be certain that
result \textquotedblleft fail\textquotedblright\ will necessarily be
obtained by $W$ only if $F$ is
allowed to go through a quantum superposition \footnote{We
have checked in \S \ \ref{F-andFbar-proj} that, if this coherent superposition
is destroyed, the possibility of result \textquotedblleft OK\textquotedblright%
\ reappears.} of $\left\vert F:\uparrow\right\rangle $ and $\left\vert
F:\downarrow\right\rangle $, of $\left\vert F:\uparrow
\right\rangle$ and $\left\vert F:\downarrow
\right\rangle$. One can save  both statements $s_Q^{\overline F}$ and $s_Q^W$, but at the
price of treating the two friends $\overline{F}$ and $F$  in a non symmetric way, and losing $s_Q^{\overline W}$.

(ii) One could also envisage a \textquotedblleft mirror\textquotedblright\ case where O allows $\overline{F}$ to be described after measurement by quantum
superpositions of $\left\vert \text{heads}\right\rangle $ and $\left\vert
\text{tails}\right\rangle $, but considers that $F$ is restricted to quantum states
associated with well-defined results. In this case, we have seen in (\ref{5}) that result
$\overline{\text{OK}}$ observed by $\overline{W}$ means that the spin is in
state $\left\vert S:\uparrow\right\rangle $; as a consequence, $W$ can observe
both results OK and \textquotedblleft fail\textquotedblright. But then
$\overline{F}$ has been projected into another coherent superposition of
$\left\vert \text{heads}\right\rangle $ and $\left\vert \text{tails}%
\right\rangle $, and cannot make any specific statement at any time.

(iii) If O allows $\overline{F}$ and $F$ to be described by any superposition after they have performed a measurement, he does not get any certain statement from $\overline{F}$ at time $t=0$, and therefore cannot write the fourth line of Table II. At later times, we have seen\footnote{Compare the third line of (\ref{6-ter}) and
(\ref{9-5}).} in \S \ \ref{correlations} that the perturbation introduced by
$\overline{W}$'s measurement destroys the perfect correlation between r=tails
and w=\textquotedblleft fail\textquotedblright. This is because the perturbation
has erased any specific result contained in $\overline{F}$'s memory, and put
her \textquotedblleft memory register\textquotedblright\ (in\ Everett's terms)
into a state containing at the same time different results; under these
conditions, one may wonder how she could express any statement with certainty
to O.

(iv) Finally, if neither of the friends $\overline{F}$ and $F$ is allowed to
remain after measurement in quantum superpositions of states associated with
different measurement results, we have seen in \S \ \ref{F-andFbar-proj} that
the successive projections destroy the correlations.

In conclusion of this discussion, if O wants to write all statements of Table II\ together, he must
accept a statement made by $\overline F$ about a specific result she has observed one specific result
even if, at the same time, $\overline F$ is actually described by a state
 vector where its memory contains several different results. This amounts to combining the properties
 arising from several state vectors, i.e. combining correlations properties associated with incompatible
 measurements ($\overline{F}$'s and\ $W$'s
measurements are associated with eigenstates that, for photons, would be at 45
degrees from each other, and therefore incompatible measurements). But consistency requires that he should use a single state vector: just after $\overline F$ performed her measurement, either she is
in an eigenstate of measurement (and she can then issue a valid statement), or she is
in a quantum superposition of states associated with several different results (she then cannot make any statement), but not both at the same time.

\section{Reversing the order of measurements}

\label{reversing}

Another possibility is to try changing the orders of the measurements, in
order to explore if this provides a way to obtain all statements of table II
at the same time.

\subsection{$\overline{W}$ performs his experiment before $\overline{F}$}

\label{W-bar-before}

Assume now that $\overline{W}$ performs his experiment before $\overline{F}$:
the order of measurements is now $F$, $\overline{W}$, and then only
$\overline{F}$.\ When $\overline{W}$ performs his measurement, operator
$\overline{F}$\ and his laboratory still in their initial state, which we
denote $\left\vert \overline{F}:0\right\rangle $; the eigenstates of $W$'s
measurement are obtained by replacing, in (\ref{4}), both kets $\left\vert
\overline{F}:h\right\rangle $ and $\left\vert \overline{F}:t\right\rangle $ by
$\left\vert \overline{F}:0\right\rangle $. When this substitution is made, the
state vector $\left\vert \Psi_{4}^{\prime\prime}\right\rangle $ after
$\overline{W}$\ has obtained result $\overline{\text{OK }}$ is obtained by
using (\ref{5}):%
\begin{equation}
\left\vert \Psi_{4}^{\prime\prime}\right\rangle =\left\vert \overline
{\text{OK}}\right\rangle _{\overline{L}}\left\vert \overline{W}:\overline
{\text{OK}}\right\rangle \left\vert S:\uparrow\right\rangle \left\vert
F:\uparrow\right\rangle =\frac{1}{\sqrt{2}}\Big[\left\vert \text{heads}%
\right\rangle -\left\vert \text{tails}\right\rangle \Big]\left\vert
\overline{F}:0\right\rangle \left\vert \overline{W}:\overline{\text{OK}%
}\right\rangle ~\left\vert S:\uparrow\right\rangle \left\vert F:\uparrow
\right\rangle \label{10-prime}%
\end{equation}
In this way, the correlation between the measurement result $\overline
{\text{OK}}$ and $\left\vert S:\uparrow\right\rangle $ is obtained again.\ But
then, if $\overline{F}$ performs his measurement and obtains result
\textquotedblleft tails\textquotedblright, the state vector becomes:%
\begin{equation}
\left\vert \Psi_{5}^{\prime\prime}\right\rangle =\left\vert \text{tails}%
\right\rangle \left\vert \overline{F}:t\right\rangle ~\left\vert \overline
{W}:\overline{\text{OK}}\right\rangle ~\left\vert S:\uparrow\right\rangle
\left\vert F:\uparrow\right\rangle =\left\vert \text{tails}\right\rangle
\left\vert \overline{F}:t\right\rangle ~\left\vert \overline{W}:\overline
{\text{OK}}\right\rangle ~\Big[\left\vert \text{fail}\right\rangle
_{L}-\left\vert \text{OK}\right\rangle _{L}\Big] \label{11-prime}%
\end{equation}
operator $\overline{F}$ can non longer be sure that observing
\textquotedblleft tails\textquotedblright\ implies that $W$ will obtain any
specific result. Reversing the order of $\overline{W}$ and $\overline{F}%
$\ experiments restores the perfect correlation between results $\overline
{\text{OK}}$ and $\left\vert S:\uparrow\right\rangle $, but cancels that
between \textquotedblleft tails\textquotedblright\ and \textquotedblleft
fail\textquotedblright.

\subsection{$W$ performs his measurement before $\overline{W}$}

\label{simpler-experiment}

In order to better conserve the correlation between the result observed by
$\overline{F}$ and that obtained by $W$, we now assume that $W$ performs his
measurement before $\overline{W}$ does (or that $W$ performs no measurement at all).

\subsubsection{Without projection the state vector}

\label{simpler-experiment-no-projection}

In this case, we need to directly expand the ket (\ref{3}) onto the
eigenstates (\ref{6}), which provides:
\begin{align}
\left\vert \Psi_{3}\right\rangle  &  =\frac{1}{\sqrt{6}}\Big[\left\vert
\text{heads}\right\rangle \left\vert \overline{F}:h\right\rangle +\left\vert
\text{tails}\right\rangle \left\vert \overline{F}:t\right\rangle
\Big]~\Big[\left\vert \text{OK}\right\rangle _{L}+\left\vert \text{fail}%
\right\rangle _{L}\Big]+\frac{1}{\sqrt{6}}\left\vert \text{tails}\right\rangle
\left\vert \overline{F}:t\right\rangle \Big[\left\vert \text{fail}%
\right\rangle _{L}-\left\vert \text{OK}\right\rangle _{L}\Big]\nonumber\\
&  =\frac{1}{\sqrt{6}}\left\vert \text{heads}\right\rangle \left\vert
\overline{F}:h\right\rangle ~\Big[\left\vert \text{OK}\right\rangle
_{L}+\left\vert \text{fail}\right\rangle _{L}\Big]+\sqrt{\frac{2}{3}%
}\left\vert \text{tails}\right\rangle \left\vert \overline{F}:t\right\rangle
\left\vert \text{fail}\right\rangle _{L} \label{10}%
\end{align}
This provides the sate vector just before $W$ performs his measurement.

In the right hand side of this result, we notice the absence of any term in
$\left\vert \text{tails}\right\rangle \left\vert \overline{F}:t\right\rangle
\left\vert \text{OK}\right\rangle _{L}$, due to a destructive interference
between two terms in the first line of (\ref{10}). Now that the perturbation
introduced by $\overline{W}$'s measurement has disappeared, we see that
$\left\vert \text{tails}\right\rangle $ and $\left\vert \text{fail}%
\right\rangle _{L}$ are indeed perfectly correlated. Statement $s_{Q}%
^{\overline F}$ in Table II of Ref. \cite{Frauchiger-Renner} then becomes
correct, but statement $s_{Q}^{\bar W}$ vanishes.

\subsubsection{With a projection the initial state vector}

\label{simpler-experiment-projection}

Another way to check that $\left\vert \text{tails}\right\rangle $ and
$\left\vert \text{fail}\right\rangle _{L}$ are perfectly correlated is to
assume that $\overline{F}$ measures \textquotedblleft tails\textquotedblright,
and applies the projection postulate to $\left\vert \Psi_{1}\right\rangle $
given by (\ref{1}).\ Then $\left\vert \Psi_{2}\right\rangle $ in (\ref{2}) is
truncated to its $\left\vert \text{tails}\right\rangle $ component (multiplied
by $\sqrt{3/2}$ for normalization) and $\left\vert \Psi_{3}\right\rangle $
becomes:%
\begin{equation}
\left\vert \Psi_{3}^{\prime}\right\rangle =\frac{1}{\sqrt{2}}\left\vert
\text{tails}\right\rangle \left\vert \overline{F}:t\right\rangle
\Big[\left\vert S:\downarrow\right\rangle \left\vert F:\downarrow\right\rangle
+\left\vert S:\uparrow\right\rangle \left\vert F:\uparrow\right\rangle
\Big]=\left\vert \text{tails}\right\rangle \left\vert \overline{F}%
:t\right\rangle \left\vert \text{fail}\right\rangle _{L} \label{11}%
\end{equation}
Clearly, $\overline{F}$ can then predict with certainly that $W$'s measurement
will give the result \textquotedblleft fail\textquotedblright.\ This is easy
to understand physically: if $\overline{F}$ obtains a tails, she sends a spin
in a transverse $Ox$ direction to $L$, so that $F$ and her laboratory $L$
reach a coherent superposition that is orthogonal to $\left\vert
\text{OK}\right\rangle _{L}$.\ Result OK is then impossible, due to a
destructive interference between the components $\left\vert S:\downarrow
\right\rangle \left\vert F:\downarrow\right\rangle $ and $\left\vert
S:\uparrow\right\rangle \left\vert F:\uparrow\right\rangle $. Note in passing
that this assumes and interference between two macroscopically distinct states
($F$ is macroscopic): the measurement apparatus used by $W$ necessarily makes
these states overlap again (in the dBB interpretation, one says that a
macroscopic wave that was empty becomes effective again).

But, if $\overline{W}$ performs his measurement before $W$, this destructive
interference effect no longer occurs: the coefficients are not the same in
(\ref{7}) and in (\ref{10}).\ If $\overline{F}$ has obtained \textquotedblleft
tails\textquotedblright, $\overline{W}$ can then obtain both results. If for
instance he obtains result $\overline{OK}$, one has to project (\ref{7}) onto
states containing $\left\vert \overline{\text{OK}}\right\rangle _{\overline
{L}}$, and one obtains (after normalization):%
\begin{align}
\left\vert \overline{\Psi}_{5}\right\rangle  &  =\frac{1}{\sqrt{2}}\left\vert
\overline{\text{OK}}\right\rangle _{\overline{L}}\left\vert \overline
{W}:\overline{\text{OK}}\right\rangle \left\vert \text{OK}\right\rangle
_{L}\left\vert W:\text{OK}\right\rangle -\frac{1}{\sqrt{2}}\left\vert
\overline{\text{OK}}\right\rangle _{\overline{L}}\left\vert \overline
{\text{OK}}\right\rangle _{\overline{L}}\left\vert \text{fail}\right\rangle
_{L}\left\vert W:\text{fail}\right\rangle \nonumber\\
&  =\left\vert \overline{\text{OK}}\right\rangle _{\overline{L}}\left\vert
\overline{W}:\overline{\text{OK}}\right\rangle \left\vert \text{OK}%
\right\rangle \left\vert S:\downarrow\right\rangle \left\vert F:\downarrow
\right\rangle \label{12}%
\end{align}
Now state $\left\vert S:\uparrow\right\rangle \left\vert F:\uparrow
\right\rangle $ has completely disappeared from this ket, so that the
destructive interference effect can non longer take place; none of the final
results, OK or fail is forbidden in the $\left\vert \text{tails}\right\rangle
$ component. This is also visible in the expression:%
\begin{equation}
\left\vert \overline{\Psi}_{5}\right\rangle =\frac{1}{\sqrt{2}}\left\vert
\overline{W}:\overline{\text{OK}}\right\rangle \left\vert \text{OK}%
\right\rangle ~\Big[\left\vert \text{OK}\right\rangle _{L}+\left\vert
\text{fail}\right\rangle _{L}\Big] \label{12-bis}%
\end{equation}

\section{Adding a qubit to the experiment}

\label{third-experiment}

It has also been suggested \cite{Grangier} to protect the information obtained
by $\overline{F}$ from the perturbations created by the outside observers, and
to assume that she uses a \textquotedblleft secret qubit\textquotedblright\ to
store her result of measurement. The hope is that this additional qubit may
connect the situations of \S ~\ref{first-expt-no-projn} and
\ref{first-expt-F-projn}, by providing a which-path information at the level
of $\bar F$ that remains perfectly stable. Alternatively, this qubit can be
seen as representing a \textquotedblleft friend of the
friend\textquotedblright, which we name $\overline{G}$. In any case, the corresponding
quantum system reaches state $\left\vert \overline{G}:h\right\rangle $ if
$\overline{F}$ observes \textquotedblleft heads\textquotedblright, and state
$\left\vert \overline{G}:t\right\rangle $ if she observes \textquotedblleft
tails\textquotedblright.\ Then $\left\vert \Psi_{3}\right\rangle $ becomes:
\begin{align}
\left\vert \widehat{\Psi}_{3}\right\rangle  &  =\frac{1}{\sqrt{3}%
}\Big[\left\vert \text{heads}\right\rangle \left\vert \overline{F}%
:h\right\rangle \left\vert \overline{G}:h\right\rangle +\left\vert
\text{tails}\right\rangle \left\vert \overline{F}:t\right\rangle \left\vert
\overline{G}:t\right\rangle \Big]\left\vert S:\downarrow\right\rangle
\left\vert F:\downarrow\right\rangle \nonumber\\
&  +\frac{1}{\sqrt{3}}\left\vert \text{tails}\right\rangle \left\vert
\overline{F}:t\right\rangle \left\vert \overline{G}:t\right\rangle \left\vert
S:\uparrow\right\rangle \left\vert F:\uparrow\right\rangle \label{14}%
\end{align}

The external operator $\overline{W}$ now performs a measurement on
$\overline{L}$ with unchanged eigenstates (\ref{4}).\ In this basis,
$\left\vert \widehat{\Psi}_{3}\right\rangle $ becomes:%
\begin{align}
\left\vert \widehat{\Psi}_{3}\right\rangle  &  =\frac{1}{\sqrt{6}}\left\{
\Big[\left\vert \overline{\text{OK}}\right\rangle _{\overline{L}}+\left\vert
\overline{\text{fail}}\right\rangle _{\overline{L}}\Big]\left\vert
\overline{G}:h\right\rangle +\Big[\left\vert \overline{\text{fail}%
}\right\rangle _{\overline{L}}-\left\vert \overline{\text{OK}}\right\rangle
_{\overline{L}}\Big]\left\vert \overline{G}:t\right\rangle \right\}
\left\vert S:\downarrow\right\rangle \left\vert F:\downarrow\right\rangle
\nonumber\\
&  ~~~~~~~~~~~~~~~~~~~~~~~~~~~~~~~~~~~~~~~~~~~+\frac{1}{\sqrt{6}%
}\Big[\left\vert \overline{\text{fail}}\right\rangle _{\overline{L}%
}-\left\vert \overline{\text{OK}}\right\rangle _{\overline{L}}\Big]\left\vert
\overline{G}:t\right\rangle \left\vert S:\uparrow\right\rangle \left\vert
F:\uparrow\right\rangle \label{15}%
\end{align}
After $\overline{W}$'s measurement, we obtain:%
\begin{align}
\left\vert \widehat{\Psi}_{4}\right\rangle  &  =\frac{1}{\sqrt{6}}\left\{
\Big[\left\vert \overline{\text{OK}}\right\rangle _{\overline{L}}\left\vert
\overline{W}:\overline{\text{OK}}\right\rangle +\left\vert \overline
{\text{fail}}\right\rangle _{\overline{L}}\left\vert \overline{W}%
:\overline{\text{fail}}\right\rangle \Big]\left\vert \overline{G}%
:h\right\rangle \right. \nonumber\\
&  ~~~~~~~~~~~~~~~~+\Big[\left\vert \overline{\text{fail}}\right\rangle
_{\overline{L}}\left\vert \overline{W}:\overline{\text{fail}}\right\rangle
\left.  -\left\vert \overline{\text{OK}}\right\rangle _{\overline{L}%
}\left\vert \overline{W}:\overline{\text{OK}}\right\rangle \Big]\left\vert
\overline{G}:t\right\rangle \right\}  \left\vert S:\downarrow\right\rangle
\left\vert F:\downarrow\right\rangle \nonumber\\
&  ~~~~~~~~~~~~~~~~~~~~~+\frac{1}{\sqrt{6}}\Big[\left\vert \overline
{\text{fail}}\right\rangle _{\overline{L}}\left\vert \overline{W}%
:\overline{\text{fail}}\right\rangle -\left\vert \overline{\text{OK}%
}\right\rangle _{\overline{L}}\left\vert \overline{W}:\overline{\text{OK}%
}\right\rangle \Big]\left\vert \overline{G}:t\right\rangle \left\vert
S:\uparrow\right\rangle \left\vert F:\uparrow\right\rangle \label{17}%
\end{align}

Finally, $W$ performs his measurements with unchanged eigenstates
(\ref{6}).\ The expansion of $\left\vert \widehat{\Psi}_{4}\right\rangle $ on
these eigenstates finally leads to the complicated expression:%
\begin{align}
\left\vert \widehat{\Psi}_{4}\right\rangle  &  =\frac{1}{\sqrt{12}}\left\{
\Big[\left\vert \overline{\text{OK}}\right\rangle _{\overline{L}}\left\vert
\overline{W}:\overline{\text{OK}}\right\rangle +\left\vert \overline
{\text{fail}}\right\rangle _{\overline{L}}\left\vert \overline{W}%
:\overline{\text{fail}}\right\rangle \Big]\left\vert \overline{G}%
:h\right\rangle \right. \nonumber\\
&  ~~~~~~~~~\left.  +\Big[\left\vert \overline{\text{fail}}\right\rangle
_{\overline{L}}\left\vert \overline{W}:\overline{\text{fail}}\right\rangle
-\left\vert \overline{\text{OK}}\right\rangle _{\overline{L}}\left\vert
\overline{W}:\overline{\text{OK}}\right\rangle \Big]\left\vert \overline
{G}:t\right\rangle \right\}  \Big[\left\vert \text{OK}\right\rangle
_{L}+\left\vert \text{fail}\right\rangle _{L}\Big]\nonumber\\
&  ~~~~~~~~~~~~+\frac{1}{\sqrt{12}}\Big[\left\vert \overline{\text{fail}%
}\right\rangle _{\overline{L}}\left\vert \overline{W}:\overline{\text{fail}%
}\right\rangle -\left\vert \overline{\text{OK}}\right\rangle _{\overline{L}%
}\left\vert \overline{W}:\overline{\text{OK}}\right\rangle \Big]\left\vert
\overline{G}:t\right\rangle \Big[\left\vert \text{fail}\right\rangle
_{L}-\left\vert \text{OK}\right\rangle _{L}\Big] \label{18}%
\end{align}
or:%
\begin{align}
\left\vert \widehat{\Psi}_{4}\right\rangle  &  =\left\vert \overline
{\text{OK}}\right\rangle _{\overline{L}}\left\vert \overline{W}:\overline
{\text{OK}}\right\rangle \left\vert \text{OK}\right\rangle _{L}\frac{1}%
{\sqrt{12}}\left\vert \overline{G}:h\right\rangle \nonumber\\
&  +\left\vert \overline{\text{OK}}\right\rangle _{\overline{L}}\left\vert
\overline{W}:\overline{\text{OK}}\right\rangle \left\vert \text{fail}%
\right\rangle _{L}\Big[\frac{1}{\sqrt{12}}\left\vert \overline{G}%
:h\right\rangle -\frac{2}{\sqrt{12}}\left\vert \overline{G}:t\right\rangle
\Big]\nonumber\\
&  +\left\vert \overline{\text{fail}}\right\rangle _{\overline{L}}\left\vert
\overline{W}:\overline{\text{fail}}\right\rangle \left\vert \text{OK}%
\right\rangle _{L}\frac{1}{\sqrt{12}}\left\vert \overline{G}:h\right\rangle
\nonumber\\
&  +\left\vert \overline{\text{fail}}\right\rangle _{\overline{L}}\left\vert
\overline{W}:\overline{\text{fail}}\right\rangle \left\vert \text{fail}%
\right\rangle _{L}\Big[\frac{1}{\sqrt{12}}\left\vert \overline{G}%
:h\right\rangle +\frac{2}{\sqrt{12}}\left\vert \overline{G}:t\right\rangle
\Big] \label{19}%
\end{align}
As expected, the state $\left\vert \overline{G}:t\right\rangle $ is correlated
with only one result of measurement performed by $W$, namely result
\textquotedblleft fail\textquotedblright.\ This is because $\left\vert
\overline{G}:t\right\rangle $ keeps a memory of a past event where
$\overline{F}$ sent a transversely polarized spin to $F$, which projects $L$
into a state that is orthogonal to $\left\vert \text{OK}\right\rangle _{L}$,
and therefore necessarily corresponds to result \textquotedblleft
fail\textquotedblright. But $\overline{\text{OK}}$ is not correlated to any
specific spin orientation or state of agent $F$. Replacing $\overline{F}$ by
$\overline{G}$ therefore does restore the validity of the first statement of
Table II of Ref.~\cite{Frauchiger-Renner}, but cancels others. Not surprisingly, we recover the same predictions as in \S~\ref{first-expt-F-projn}, since the selection of one quantum state of the added qubit is equivalent to a projection of the state vector. Similarly, if another hidden qubit was introduced to store $F$'s result, we would recover the results of \S~\ref{F-andFbar-proj}.

\section{Conclusion}

\label{discussion}\label{conclusion}

Because of the \textquotedblleft shifty split\textquotedblright\ problem, most
authors agree that standard quantum mechanics is, to some extent,
fundamentally an ill defined theory -- even if, in practice, physicists know
how to use it perfectly well.\ This is the reason why we have attempted to
recover all statements of Table II of Ref. \cite{Frauchiger-Renner} by
applying the projection postulate in different ways (or apply no projection at
all); we have also tried to reverse the order of experiments in order to
change the correlations.\ The conclusion is that there seems to be no way to
obtain all statements of Table II together with a single state vector;
depending on the method of calculation, some hold and some do not, but no
calculation leads to all statements at the same time. For instance, they require that a statement made by $\overline F$ after having observed a specific result should be considered as valid even if, at the same time, $\overline F$ is described by a state vector where its memory contains several different results. This amounts to combining conclusions obtained by putting the \textquotedblleft shifty split\textquotedblright\ at completely different places. Bohr would probably have seen
this situation as one more illustration of complementarity: statements that
are relative to incompatible experimental descriptions or devices vices (or different time
evolutions) should not be combined together. This can also be seen, for
instance, in Ref.~\cite{Bub-2017}, which explicitly gives the three
expressions of the state vector expanded on three different basis associated
with three incompatible couple of measurements.\ So, even if ill defined to
some extent, standard quantum mechanics remains consistent.

The reasoning of Ref. \cite{Frauchiger-Renner} requires combining the
observations and deductions of several agents, who use assumption
(Q). Nevertheless this assumption is ambiguous, and actually can be understood in different ways.\ It
may indeed be seen as an invitation for each agent to make predictions by
using his own state vector (or density matrix), and by calculating the average value of projectors
that correspond only to the measurements she/he has made (or might have
made).\ Each agent would then take into account only the specific result of her/his
measurement by projecting his state vector (or density matrix) independently of the others, and ignoring 
the effects of other measurements in the past of the
future.\ For instance, as we have seen in \S \S ~\ref{premiere-discussion} and
\ref{W-bar-before}, $\overline{W}$ can predict that his result $\overline{OK}$
necessarily corresponds to a spin $\left\vert \uparrow\right\rangle $ if he
ignores the effects of $\overline{F}$'s measurement, and takes into account
only his own projector.

But, in standard quantum mechanics (independently of
its interpretation), this is not the way a probability (or a certainty) should
be calculated.\ If several measurements are performed, the probability must be
calculated by calculating the average value of products of symmetrically
embedded projectors (this is sometimes called the Wigner probability
formula).\ For instance, if $\overline{F}$ makes the first  measurement, and if
the information on the result leaks outside (to the superoperator), the corresponding projector must be
applied first to the state vector; the discussion of the hidden qubit in \S \
\ref{third-experiment} also illustrates the role of this information.\ These successive
projections automatically prevents the profusion of different state
vectors.\ Generally speaking, if one remains within standard quantum theory,
one should not combine the predictions obtained from two different quantum
states, even if they are states describing the same physical system at
different times (for instance $\left\vert \Psi_{3}\right\rangle $ and
$\left\vert \Psi_{5}\right\rangle $ before and after $\overline{W}$ performed
his measurement). In other words, if a physical system has at a given time
certain correlation properties, there is no reason in standard theory why
these properties should remain at all times, when these correlations might
have been perturbed. Nothing of course forbids changing the rules of quantum
mechanics to invent a new theory\footnote{The modal
interpretation of quantum mechanics introduces several states vectors to
describe the same quantum system, but in a way that is carefully designed not
to change the predictions of quantum mechanics.}, but the
inconsistencies that may then be obtained should not be ascribed to quantum mechanics. In his discussion of the same subject \cite{Sudbery-2017}, A. Sudbury comes to very similar conclusions, and analyzes it in terms of the BBB theory (Bell version of the de Broglie-Bohm field theory).

Even if one decides to give up the standard rules, it is not clear how
one could reason on experiments where observers  project each other into different quantum states, mutually changing the contents of their memories.  Even if the agent had in mind a specific result she observed before, she may be put into another state where her memory now contains different measurement results at the same time, a state where they no longer know what she had observed. Once the
experiment is completed, all the observers are supposed to exchange
information and to build a reasoning together, or to send the information to the superoperator O. But what is the meaning of a
discussion between observers that are sorts of Schr\"{o}dinger cats,
remembering opposite facts at the same time? It seems more reasonable to
consider an agent as a reliable source of information only if his/her memory
contains a result that is certain.\ One could of
course always invoke a \textquotedblleft freedom of choice of the
agents\textquotedblright, and consider that the memories of each agent belong
each to their own world, which is described by their own private state vector;
but the agents then belong to parallel, incompatible, worlds, so that their
observations should not be combined in a single reasoning. Mathematically,
certainties are a special case of probabilities, and it is known that
combining them requires that they should all belong to the same space of
probabilities.\ The exact nature of this space should be defined precisely
before building a logical scheme.

Our conclusion is that quantum mechanics may be ill defined, but that the
argument of Ref. \cite{Frauchiger-Renner} does not particularly point to any
specific possible internal inconsistency. Rather, it shows that logical
inconsistencies may appear if one changes the rules by describing of the same
physical evolution with several different independent state vectors, and then
combines the corresponding predictions. Since none of the interpretations discussed in Ref.~\cite{Frauchiger-Renner}
includes this possibility, none of them leads to all 4 statements of Table II,
which means that they do not necessarily have to violate one of the 3
assumptions (Q),(C) and (S) of this reference. Of course, examining their
compatibility with the different interpretations of quantum mechanics is an
interesting question in itself.\ Nevertheless, it would also be interesting to
clarify precisely which sort of modified quantum theory could lead to all
conclusions of Table II.

\vspace{0.5cm} \textbf{Acknowledgments} \vspace{3mm}

The author is very grateful to Philippe Grangier for drawing his attention to
Ref. \cite{Frauchiger-Renner}, as well as for many interesting discussions and
useful comments.



\begin{center}
Appendix
\end{center}

It is instructive to investigate how some of the interpretations of quantum
mechanics describe the succession of events taking place in the Frauchiger-Renner experiment.\ We will examine two of them: modified Schr\"{o}dinger dynamics
\cite{Ghirardi-GRW,  Pearle-1976-1979} and dBB (de Broglie-Bohm) theory
\cite{De-Broglie, Bohm}.

($i$) Modified Schr\"{o}dinger dynamics introduces a stochastic reduction
process of the state vector. Any macroscopic superposition of
states localized in different regions of space is quickly resolved into one of its components by this process. Since
every agent is indeed macroscopic, state vector reduction occurs at each stage of
measurement, so that the evolution of the state vector is as studied in \S \
\ref{first-expt-every-agent}. All interference effects that were at the origin
of the statements of Table II of Ref.~\cite{Frauchiger-Renner} then
disappear, and the statements are not valid; no paradox occurs.

($ii$) In dBB theory, the state vector is never reduced. If the wave function
divides into several components in the configuration space, the Bohmian position follows one of them in this space, while the other components become  \textquotedblleft empty waves\textquotedblright . In a single realization of the experiment,  the Bohmian positions associated with the pointer of each measurement apparatus constantly keep a well-defined position, even if the quantum state is a superposition of different positions; we never have situations where, as for instance in the discussion at the end of \S~\ref{first-expt-F-projn}, $\overline{F}$'s pointer (or memory) indicates several different results at the same time. It is thefefore worth studying the dBB description of the experiment.

An important feature is that the Bohmian successive positions of the same pointer may indicate results that are opposite at different times (namely, before and after the measurement apparatus has been subjected to en external coherent quantum measurement). To understand how this can happen, we first assume that the initial quantum state is (\ref{1-prime}), which has been prepared by $\overline{F}$ obtaining \textquotedblleft tails\textquotedblright . After $F$'s measurement, the
Bohmian positions of $\overline{F}$ and her laboratory, of $F$ and her
laboratory, and that of the spin particle, are such that the point in
configuration space lies in a region where one of the components of
(\ref{4-prime}) does not vanish. The pointer of $\overline{F}$'s measurement apparatus is then necessarily
at one place where it indicates \textquotedblleft tails\textquotedblright , while
the pointer of $F$' measurement apparatus indicates one of the two possible spin directions.

Then comes $\overline{W}$'s measurement, which strongly perturbs the state
of $\overline{F}$ and her laboratory. After this measurement, the point of the
configuration space must still lie in a region where the wave function of the whole
system does not vanish.\ After $W$'s measurement, the corresponding ket $\left\vert \Psi_{5}^{\prime
}\right\rangle $ is readily obtained from the ket  $\left\vert \Psi
_{4}^{\prime}\right\rangle $ written in (\ref{5-prime}) and (\ref{5-ter}):%
\begin{align}
\left\vert \Psi_{4}^{\prime}\right\rangle  & =\frac{1}{\sqrt{2}}%
\Big[\left\vert \overline{\text{fail}}\right\rangle _{\overline{L}}\left\vert
\overline{W}:\overline{\text{fail}}\right\rangle -\left\vert \overline
{\text{OK}}\right\rangle _{\overline{L}}\left\vert \overline{W}:\overline
{\text{OK}}\right\rangle \Big]~\left\vert \text{fail}\right\rangle
_{L}\left\vert W:\text{fail}\right\rangle \nonumber\\
& =\frac{1}{2}\Big[\left( \vphantom{\frac{1}{2}} \left\vert \text{heads}\right\rangle \left\vert
\overline{F}:h\right\rangle +\left\vert \text{tails}\right\rangle \left\vert
\overline{F}:t\right\rangle \right)  \left\vert \overline{W}:\overline
{\text{fail}}\right\rangle \nonumber\\
& ~~~~~~~~~~~~~~-\left( \vphantom{\frac{1}{2}} \left\vert \text{heads}\right\rangle \left\vert
\overline{F}:h\right\rangle -\left\vert \text{tails}\right\rangle \left\vert
\overline{F}:t\right\rangle \right)  \left\vert \overline{W}:\overline
{\text{OK}}\right\rangle \Big]~\left\vert \text{fail}\right\rangle
_{L}\left\vert W:\text{fail}\right\rangle \label{appendix-eq-1}%
\end{align}
The Bohmian position of the pointer of $\overline{W}$'s apparatus may be in
two regions of space, indicating one of the two possible results; as in \S~\ref{first-expt-F-projn}, result OK is not correlated only with a $\left\vert
S:\uparrow\right\rangle $ spin state, so that the third statement $s_{Q}^{\overline{W}}$ of Table II vanishes.\ Nevertheless, the pointer of
$W$'s measurement apparatus is necessarily in a region of space where it
indicates a result \textquotedblleft fail\textquotedblright, in agreement
with  the first statement $s_{Q}^{\overline{F}}$.\ We see that the pointer of $F$'s measurement apparatus may have changed its
position under the effect of $\overline{W}$'s measurement; it may now
indicate \textquotedblleft heads\textquotedblright , while it initially
indicated \textquotedblleft tails\textquotedblright.\ This is nothing but a direct consequence of
the re-appearance of the $\left\vert \text{heads}\right\rangle $ component
discussed in (ii) of \S \ \ref{first-expt-F-projn}.

This curious property can be understood in terms of the existence of the
so called \textquotedblleft surrealistic Bohmian
trajectories\textquotedblright\ \cite{Englert-1992}.\ The effect of
$\overline{W}$'s measurement on $\overline{F}$'s laboratory is to recombine
two components of the wave function associated with different results obtained before, heads or tails.\ What happens is
is similar to the recombination of the two components of the wave
function studied in \cite{Tastevin-Laloe}, where the interference between these components may reverse the motion of a pointer.
In the dBB view, the pointers constantly indicate a specific
result by their position, but this indication may change in time if the whole
apparatus is subject to the perturbation of an external measurement. So, when
speaking of the indication of pointers and results of measurement, one should
carefully specify at which stage of the experiments these indications are taken into account, as
already mentioned above.

Finally, let us come back to the full state vector of  \S~\ref{first-expt-no-projn}, without initial
projection resulting from $\overline F$'s measurement. She observes \textquotedblleft
tails\textquotedblright\ if the Bohmian position of her laboratory falls in the region of
configuration space where the $\left\vert \text{tails}\right\rangle$ component of
the ket (\ref{2}) does not vanish; the $\left\vert
\text{heads}\right\rangle$ component is then empty. Moreover, since only this empty component can lead to $W$ obtaining
result OK in the future, it seems that this result should be impossible. But we have seen that
$\overline W$'s measurement may bring back the Bohmian position of $\overline F$'s
laboratory into the region of configuration space associated with this component, which therefore becomes
active again: result OK can then be observed. In the Bohmian description, it is the effect of
$\overline W$'s measurement that reactivates a wave that was empty and restores the possibility of
obtaining result OK. Agent $\overline F$ should therefore not conclude that OK is impossible if she obtains \textquotedblleft
tails\textquotedblright , because this would amount to ignoring the perturbations induced by $\overline W$'s measurement.

In conclusion, as expected, none of these interpretations leads to all statements of Table
II of Ref.~\cite{Frauchiger-Renner}.\ Therefore, their consistency does not require that
they should necessarily reject one of the three assumptions (Q), (C) and (S)
of this reference --  even if of course this does not mean either that they
should satisfy all of them.


\end{document}